\documentclass{elsart}
\usepackage{graphicx}
\usepackage{amsmath}
\usepackage{amssymb}

\begin{document}

\begin{frontmatter}

\title{Activity spectrum from waiting-time distribution}

\author[address1,thank1]{Mauro Politi}
and
\author[address2,thank1]{Enrico Scalas}

\address[address1]{Department of Physics, Universit\`a degli Studi di Milano, via Celoria 16, 20133 Milano, Italy}

\address[address2]{Department of Advanced Science and Technology, Universit\`a degli Studi del Piemonte Orientale, 15100 Alessandria, Italy}

\thanks[thank1]{
Corresponding authors \\
E-mail: mauro.politi@unimi.it, enrico.scalas@mfn.unipmn.it}

\begin{abstract}
In high frequency financial data not only returns but also waiting
times between trades are random variables. In this work, we analyze
the spectra of the waiting-time processes for tick-by-tick trades. The numerical problem, strictly related with
the real inversion of Laplace transforms, is analyzed by using
Tikhonov's regularization method. We also analyze these spectra by
a rough method using a comb of Dirac's delta functions.
\end{abstract}

\end{frontmatter}

\section{Introduction}
It has been previously shown that waiting times between orders as well as
trades do not follow an exponential distribution \cite{engle98,anowaiting,scalas06}. This phenomenon can be explained by variable activity during the trading day, leading to a suitable mixture
of exponential distributions in order to describe the distribution
of durations \cite{scalas06a}. In this paper, we study the activity spectrum,
by numerically inverting the empirical survival function. The paper is
organized as follows: In section 2, we give the basic theoretical background and we present two methods to derive the activity spectrum. In section 3, the methods are applied to real financial data. Finally, section 4 contains our conclusions.

\section{Theory}
 
In tick-by-tick financial data, the waiting time (duration), $\tau$, between
two consecutive trades is a random variable. Let us call
$\psi(\tau)$ the probability density of durations. If we suppose that
the duration process is a mixture of exponential processes, we can write:
\begin{equation}
\psi(\tau) = \int_{0}^{\infty} g(\lambda)\lambda e^{-\lambda \tau}
d\lambda\,, \label{eq:fredholm1}
\end{equation}
where $g(\lambda)$ is the {\em spectrum of activity} satisfying
\begin{equation}
\int_0^{\infty} g(\lambda) d\lambda=1\,.
\end{equation} 
A similar equation
can be written for the {\em survival function} (the complementary
cumulative distribution function) 
$\Psi(\tau) = 1 -\int_{0}^{\tau} \psi(\tau') \, d \tau'$:
\begin{equation}
\Psi(\tau) = \int_{0}^{\infty} g(\lambda) e^{-\lambda \tau}
d\lambda\, \label{eq:fredholm2}.
\end{equation}

From eq. (\ref{eq:fredholm2}), the activity spectrum can be seen as the
solution of a Fredholm problem of the first kind:
\begin{equation}
\Psi(\tau) = \int_{0}^{\infty} g(\lambda)K(\lambda,\tau)
d\lambda\,, \label{eq:fredholm}
\end{equation}
with the kernel $K$ equal to
\begin{equation}
K(\lambda,\tau) =  e^{-\lambda \tau}\,.
\end{equation}

$\Psi(\tau)$ is indeed easily accessible from empirical data; however
the problem becomes the real inversion of a Laplace
transform for discrete and noisy real data \cite{lap1}. 

We can rewrite our linear problem, defining a matrix
\begin{equation}
K=\{k_{ij}\} \,\,\,\, k_{ij} = e^{-hij}\,, \label{eq:matrix}
\end{equation}
for $i,j = 1,...,\tau_{max}$, where $\tau_{max}$ is the largest
waiting time; the value of the $h$-parameter has to be chosen
with the aim of covering all the spectrum. In fact, we can think of
the index $j$ as the waiting time, whereas $\lambda_i =h i$ are
values of $\lambda$ in which we want to determine the unknown function
$g(\lambda)$. The equation \ref{eq:fredholm2} then becomes a matrix equation:
\begin{equation}
{\bf \Psi} = K {\bf g}\,,
\end{equation}
where $\bf{\Psi}$ is the vector of $\Psi$ values and ${\bf g}$ is the
unknown vector of activities $g(\lambda_i)$.
The problem is ill-conditioned, in fact
the ratio between the maximum and minimum elements in the matrix $K$ is
equal to $e^{h(\tau_{max}^2 - 1)}$ and the ratio between the maximum and
minimum empirical values of $\Psi$ is equal to the number of 
duration data points. For the data set described below, $\tau_{max} = 196 s$
and the number of durations is $55559$.

\subsection{Tikhonov's Method} 

Several techniques are available in
applied mathematics to solve ill-conditioned linear systems. One of
the most powerful and commonly used is Tikhonov's
regularization method \cite{tikho,hofmann,gro}.

We can think of the solution of a linear system as the
minimum of the functional
\begin{equation}
L[{\bf g}]=||K{\bf g}-\bf{\Psi}||^2\,,
\end{equation}
and the key idea of the method is to introduce a regularization
parameter, a positive real number $\mu$, and a regularization matrix (often the identity matrix $I$) such that the functional becomes
\begin{equation}
\hat{L}_{\mu}[\hat{\bf g}]=||K\hat{\bf g}_\mu-{\bf \Psi}||^2 + \mu
|| {\hat{\bf g}_\mu}||^2\,\,\,\, \mu>0.
\end{equation}
Some theorems are available for error estimation \cite{tikho,hofmann,gro}.

The procedure to find the minimum of $\hat{L}_{\mu}[\hat{\bf x}]$
can be reduced to the problem of determining an inverse
matrix for an optimal value of $\mu$. In fact, one can
show that
\begin{equation} \hat{\bf g}_\mu = (K^T K + \mu I)^{-1} K^T
\bf{\Psi}\,,
\end{equation}
where $K^T$ is the transpose of $K$.
In order to find the optimal value of $\mu$ it is usually possible to apply
the Generalized Cross Validation technique \cite{GCV} or the L-curve method
\cite{l-curve}, that are less subjected to ill-conditioning of the
matrix, but in our case, the matrix is too ill-conditioned even for these methods.
To circumvent this difficulty, we have used
Tikhonov's method for a large number of different $\mu$ values and we have
compared the rebuilt survival function
\begin{equation} \
\hat{\bf \Psi}_\mu=K\hat{\bf g}_\mu
\end{equation}
with the empirical one by means of the Kolmogorov-Smirnov test. The best
result will be the best fit of the empirical data.

\subsection{The method of Dirac's delta comb} 

In this case, we assume the
spectrum to be a {\em comb} of Dirac's delta functions:
\begin{equation}
g(\lambda)=\sum_{i=1}^M a_i \delta(\lambda-\lambda_i)\,,
\label{comb}
\end{equation}
where $M$ is a suitable number of time intervals of constant activity, $\lambda_i$, in which the trading
period has been divided and $a_i$ are suitable weights such that $\sum_{i=1}^M a_i =1$.
As a conseguence, the survival function becomes:
\begin{equation}
\Psi(\tau)=\sum_{i=1}^M a_i e^{-\lambda_i
\tau}\,.
\end{equation}
We use the following procedure to estimate the parameters $a_i$ and $\lambda_i$.
Let us fix a time-window, $\Delta T$, and let us consider the minimum number
$N_j$ of waiting times for which the sum
\begin{equation}
T_j = \sum_{i=1}^{N_j} \tau_i\,
\end{equation}
is larger than $\Delta T$. Then a term is added in eq. (\ref{comb})
with the following parameters:
\begin{equation}
\lambda_j = N_j / T_j\,\,\,\,\,\, a_j=N_j/N\,.
\end{equation}
where $N$ is the total number of data. The new interval starts
when $\tau_{N_j}$ occurs. In this way the
normalization
\begin{equation}
\sum_{i=1}^{M} a_i =1
\end{equation}
arises naturally. With this method,
the value $M$ is unknown at the beginning. Again, we test the
rebuilt survival function with the Kolmogorov-Smirnov test for different
values of $\Delta T$ in order to find the optimal size. We used this simple
method also to estimate the
parameter $h$ in the matrix (\ref{eq:matrix}).

\begin{figure}
\begin{center}
\includegraphics[height=8cm]{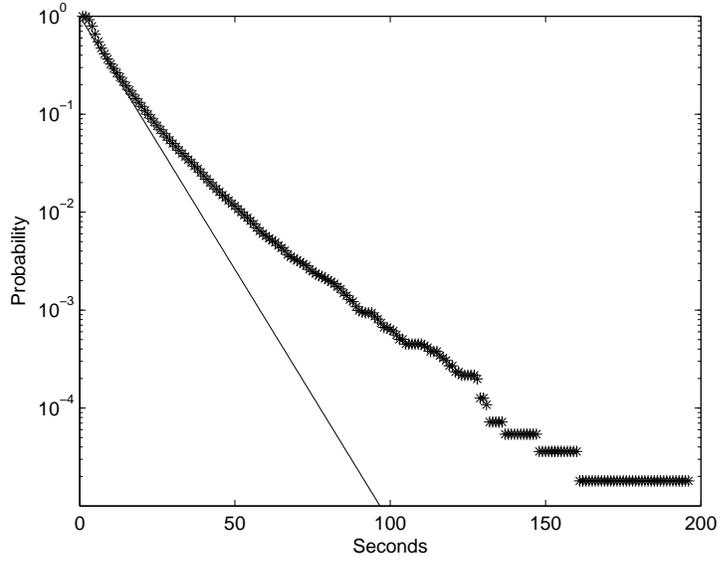}
\label{fig:survivor} \caption{Survival function for GE OCT 99 data.
The solid line represent an exponential fit with $\lambda= \frac{1}{\tau_0} $}
\end{center}
\end{figure}

\begin{figure}
\begin{center}
\includegraphics[height=8cm]{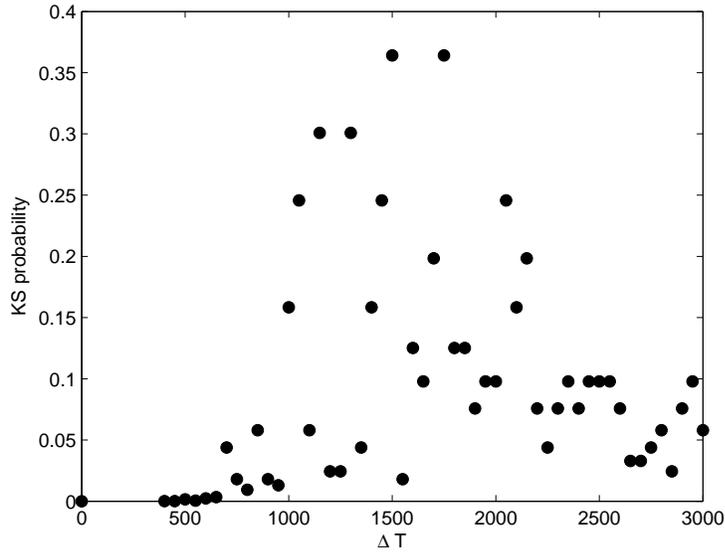}
\label{fig:timege} \caption{General Electric data analysis. KS
probability as a function of the time interval.}
\end{center}
\end{figure}

\begin{figure}
\begin{center}
\includegraphics[height=8cm]{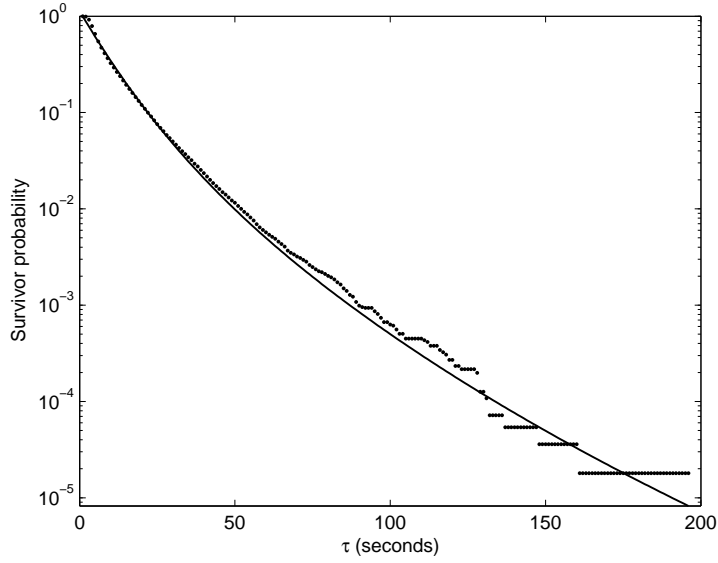}
\label{fig:timesurge} \caption{Dotted line: empirical survival
function. Solid line: survival function built by means of the time-splitting
method with $\Delta T=1500s$.}
\end{center}
\end{figure}

\begin{figure}
\begin{center}
\includegraphics[height=8cm]{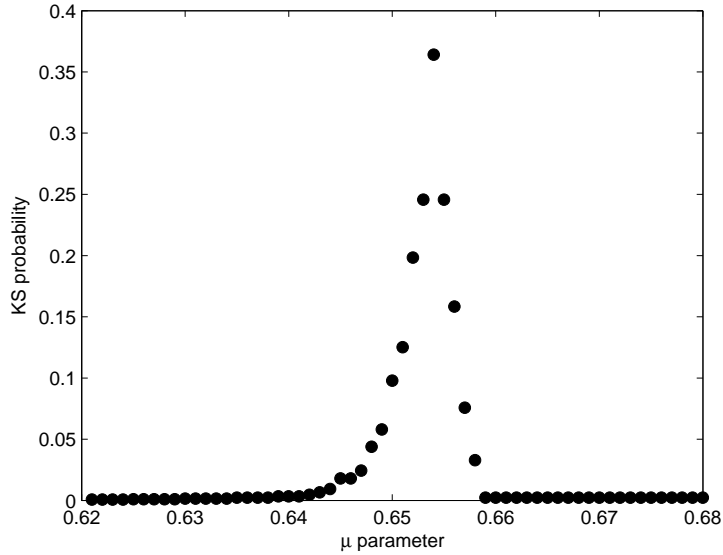}
\label{fig:tisurprob} \caption{KS probability vs $\mu$ for the
Tikhonov method. We can see that the optimal value of $\mu$ is around
$0.654$}
\end{center}
\end{figure}

\begin{figure}
\begin{center}
\includegraphics[height=8cm]{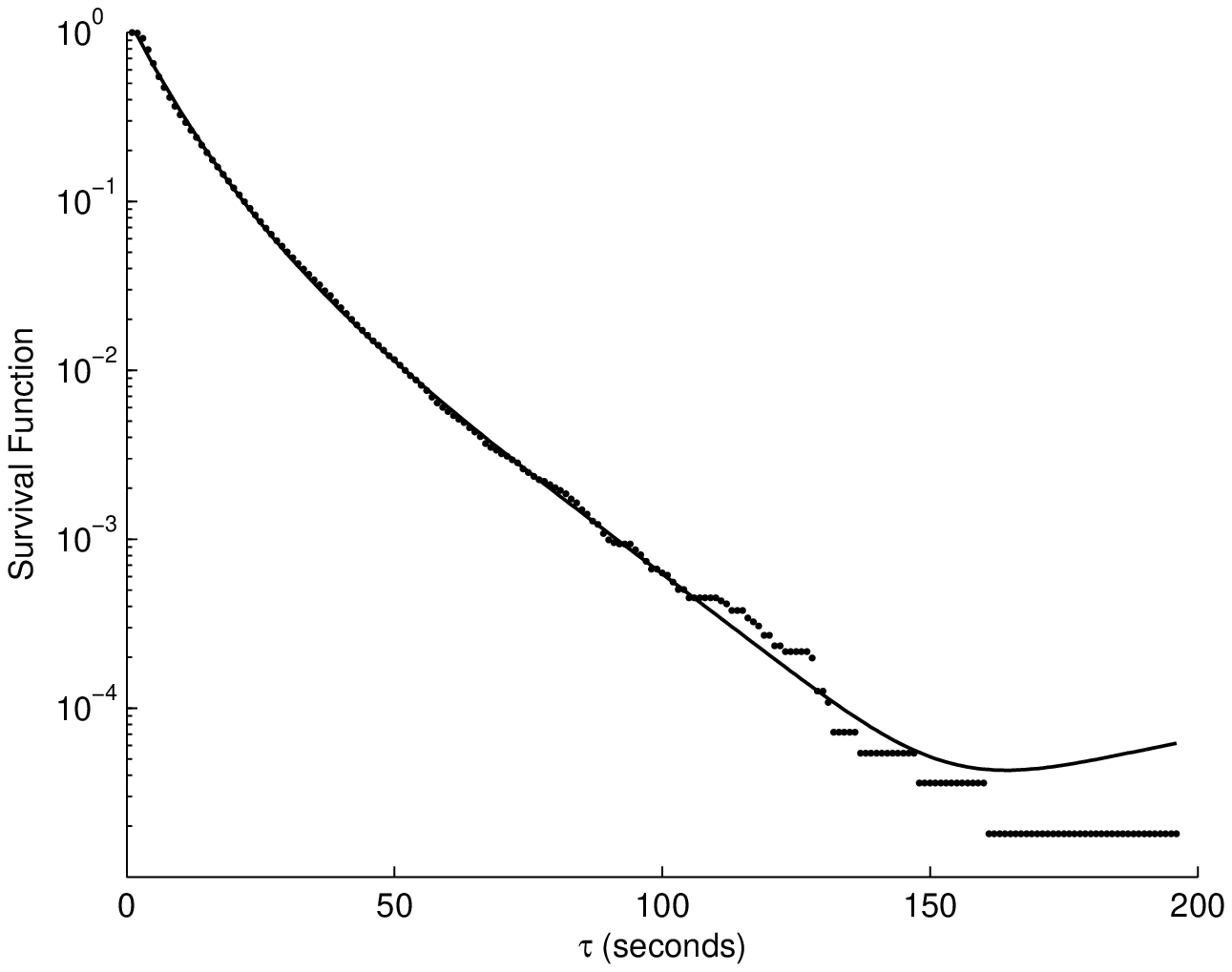}
\label{fig:tisur} \caption{Dotted line: empirical survival function.
Solid line: survival function rebuilt using the Tikhonov method
with $\mu = 0.654$.}
\end{center}
\end{figure}

\section{Results} 

In order to calibrate the methods, we have first applied them to synthetic data sets extracted from and exponential or a Mittag-Leffler distribution
\cite{ML,MLrand}. Here, however, we only report results on activity spectrum estimates for real market data.

As in ref.\cite{anowaiting}, we have considered NYSE General Electric
tick-by-tick data of October 1999. After filtering the data, 55559 waiting times were recorded,
with a mean $\tau_0 \simeq 8.85 s$. The empirical
survival function is shown in figure Fig.1.

The matrix in eq. (\ref{eq:matrix}) has a
free parameter: $h$. It defines the range of $\lambda$ in which we
can evaluate the function $g(\lambda)$. In our
case, we fix $h$ to be $0.0015$ based on a preliminary analysis
using the Dirac's delta comb method. 
The size of the matrix is $\tau_{max} \times \tau_{max} = 
196 \times 196$. Therefore, our spectrum ranges from $0$ to 
$\sim 2.5$  times $\lambda_0=1/\tau_0$.

As for the time-split method, we found that the optimal value for $\Delta T$ is around $1500 s$ (see Fig. 2). In
Fig. 3, we present the rebuilt survival function.

As for Tikhonov's method, in Fig. 4, we present the
goodness of fit based on the KS test as a function of the parameter $\mu$. Using this criterion, the
optimal value is $\mu \simeq 0.654$. In Fig. 5, we can see
the optimal reconstruction of the empirical survival function.

\section{Summary and conclusions}

By assuming that the survival function of intertrade durations can be 
written as a mixture of exponential distributions, we have proposed two methods to reconstruct the activity spectrum. The first method is based on Tikhonov's regularization. The second method uses an {\em ansatz} of intervals of constant activity. In this paper, we have not given any rigorous convergence proof and the methods outlined above were just heuristic.

The code used for this paper is available from \cite{site} or it can be obtained from the authors. Unfortunately, for copyright reasons, we cannot publish the market data-set, but we can provide a full synthetic data-set based on the Mittag-Leffler function. More details about the methods will be available in a forthcoming paper as well as in the PhD thesis by Mauro Politi.

\section*{Acknowledgments}

E.S. is grateful to Daniel Fulger and Guido Germano for pointing him
to the paper by T.J. Kozubowski and S.T. Ratchev. This activity has been
partially supported by a grant provided by East Piedmont University
(Universit\`a del Piemonte Orientale).

\end{document}